\documentclass[9pt,shortpaper,twoside,web]{ieeecolor}
\usepackage{generic}
\usepackage{cite}
\usepackage{amsmath,amssymb,amsfonts}
\usepackage{algorithmic}
\usepackage{graphicx}
\usepackage{textcomp}
\usepackage{color}
\usepackage{siunitx}
\usepackage{upgreek}
\usepackage{url}
\usepackage{mathtools}
\usepackage{soul}
\makeatletter
\let\NAT@parse\undefined
\makeatother
\usepackage{hyperref}
\hypersetup{
	colorlinks=true,
	linkcolor=blue,  
	urlcolor=blue,
	citecolor=blue
}
\def\BibTeX{{\rm B\kern-.05em{\sc i\kern-.025em b}\kern-.08em
    T\kern-.1667em\lower.7ex\hbox{E}\kern-.125emX}}
\begin{document}

\title{Inflection Phenomenon in Cryogenic MOSFET Behavior}
\author{Arnout Beckers, Farzan Jazaeri, and Christian Enz, \IEEEmembership{Fellow, IEEE}
\thanks{This project has received funding from the European Union's Horizon 2020 Research \& Innovation Programme under grant agreement No. 688539 MOS-Quito (MOS-based Quantum Information Technology).}
\thanks{The authors are with the Integrated Circuits Laboratory (ICLAB) at the Ecole Polytechnique F\'ed\'erale de Lausanne (EPFL), Microcity, 2000 Neuch\^atel, Switzerland (e-mail: arnout.beckers@epfl.ch).}}
\maketitle
\begin{abstract}
This brief reports analytical modeling and measurements of the inflection in MOSFET transfer characteristics at cryogenic temperatures. Inflection is the inward bending of the drain current versus gate voltage, which reduces the current in weak and moderate inversion at a given gate voltage compared to the drift-diffusion current. This phenomenon is explained by introducing a Gaussian distribution of localized states centered around the band edge. The localized states are attributed to disorder and interface traps. The proposed model allows to extract the density of localized states at the interface from dc current measurements. 
\end{abstract}


\section{Introduction}
\label{sec:introduction}
Cost-effective cryo-CMOS design requires characterizing and modeling of FETs at cryogenic temperatures \cite{reilly_engineering_2015,bohuslavskyi_28nm_2017,beckers_essderc,homulle_cryogenic_2018,incandela_characterization_2018,esfeh,pascal,solidstate,tedpaper}. As a side benefit of these efforts, low-temperature studies usually lead to a deeper understanding of the degradation phenomena. In 1982, Kamgar described two degraded features observed in the MOSFET's transfer characteristics at cryogenic temperatures: (i) saturation of the subthreshold swing ($SS$) with decreasing temperature ($T$), staying well above the Boltzmann limit $(k_BT/q)\ln10$, and (ii) increased inward bending (inflection) of the current that starts at a current level significantly lower than at room temperature (see Fig.\,1 in \cite{kamgar}). Kamgar's first observation, the $SS$ saturation, has recently been elucidated. At deep-cryogenic temperatures, the characteristic decay of an exponential band tail in the bandgap ($W_t$) becomes larger than the thermal energy, and the Boltzmann limit needs to be replaced by $(W_t/q)\ln 10$ \cite{bohus,phdthesisheorhii,edl}. This band tail is attributed to intrinsic mechanisms, e.g., electron-electron interactions, electron-phonon scattering, etc.\cite{sarang}. The saturation of $SS$ is universally observed in MOSFETs (see Fig.\,1 in \cite{edl}). In this work, we focus on Kamgar's second observation, the inflection phenomenon on top of the saturation of $SS$. A cryogenic model without inflection would overestimate the current in the interesting regions for power savings in qubit control electronics. Bohuslavskyi et al. reproduced the current with inflection over more than six decades of current by using an ad-hoc exponential energy dependence in the slope factor \cite{bohus,phdthesisheorhii}. Two characteristic decays of the exponential band tail were introduced without a complete physical justification to explain both the $SS$ saturation and the inflection [see Fig.\,3(c) in \cite{bohus}]. Here, we propose an alternative explanation with accompanying analytical model. The main idea is to introduce a Gaussian distribution of localized states to model the inflection, and to keep the exponential tail of the conduction band only for mobile states for $SS$ saturation.

\section{\label{sec:exp}Experimental Results}
\begin{figure*}[t]
	\centering
	\includegraphics[width=0.95\textwidth]{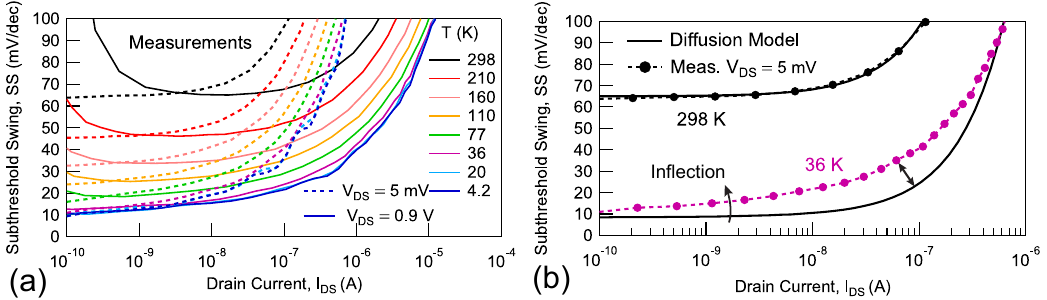}
	\vspace{-0.2cm}
	\caption{a) Subthreshold swing obtained from measured transfer characteristics, b) Standard diffusion model shows a discrepancy with the measured $SS$ at cryogenic temperatures due to the inflection phenomenon. Arrows indicate the inflection phenomenon and the decreased curvature around the knee.}
	\label{fig:meas}
\end{figure*}

The transfer characteristics of a large, $n$-channel MOSFET fabricated in a commercial 28-nm bulk CMOS process were recorded at temperatures down to \SI{4.2}{\kelvin} using a Lakeshore CPX cryogenic probe station and a Keysight B1500a parameter analyzer. Figure \ref{fig:meas}(a) shows the $SS$ extracted from these measurements and plotted versus the drain current ($I_{DS}$) at low and high drain-source voltages ($V_{DS}$). By comparing the measurements with the standard diffusion model at 298\,K and 36\,K in Fig.\,\ref{fig:meas}(b), the inflection phenomenon is clear (indicated by a flexed arrow). The inflection (i) reduces the current in weak and moderate inversion, (ii) degrades $SS$, and (iii) decreases the curvature around the knee (indicated by a double arrow). The phenomenon is present at low and high $V_{DS}$ and is essentially the same as in a 28-nm FDSOI process\cite{bohus,phdthesisheorhii} and earlier nodes\cite{kamgar}.

\begin{figure}[b]
	\centering
	\includegraphics[width=0.45\textwidth]{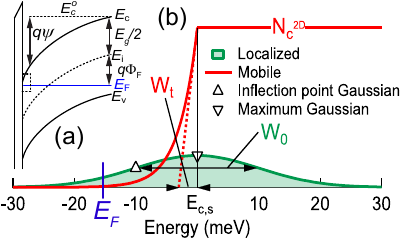}
	\vspace{-0.2cm}
	\caption{a) MOSFET band diagram, b) Zoom-in on the conduction band  at the surface. Band broadening (red, mobile states) gives $SS$ saturation \cite{bohus,edl}. Gaussian distribution (green, localized states) gives inflection.}
	\label{fig:band}
\end{figure}
\begin{figure*}[t]
	\centering
	\includegraphics[width=0.85\textwidth]{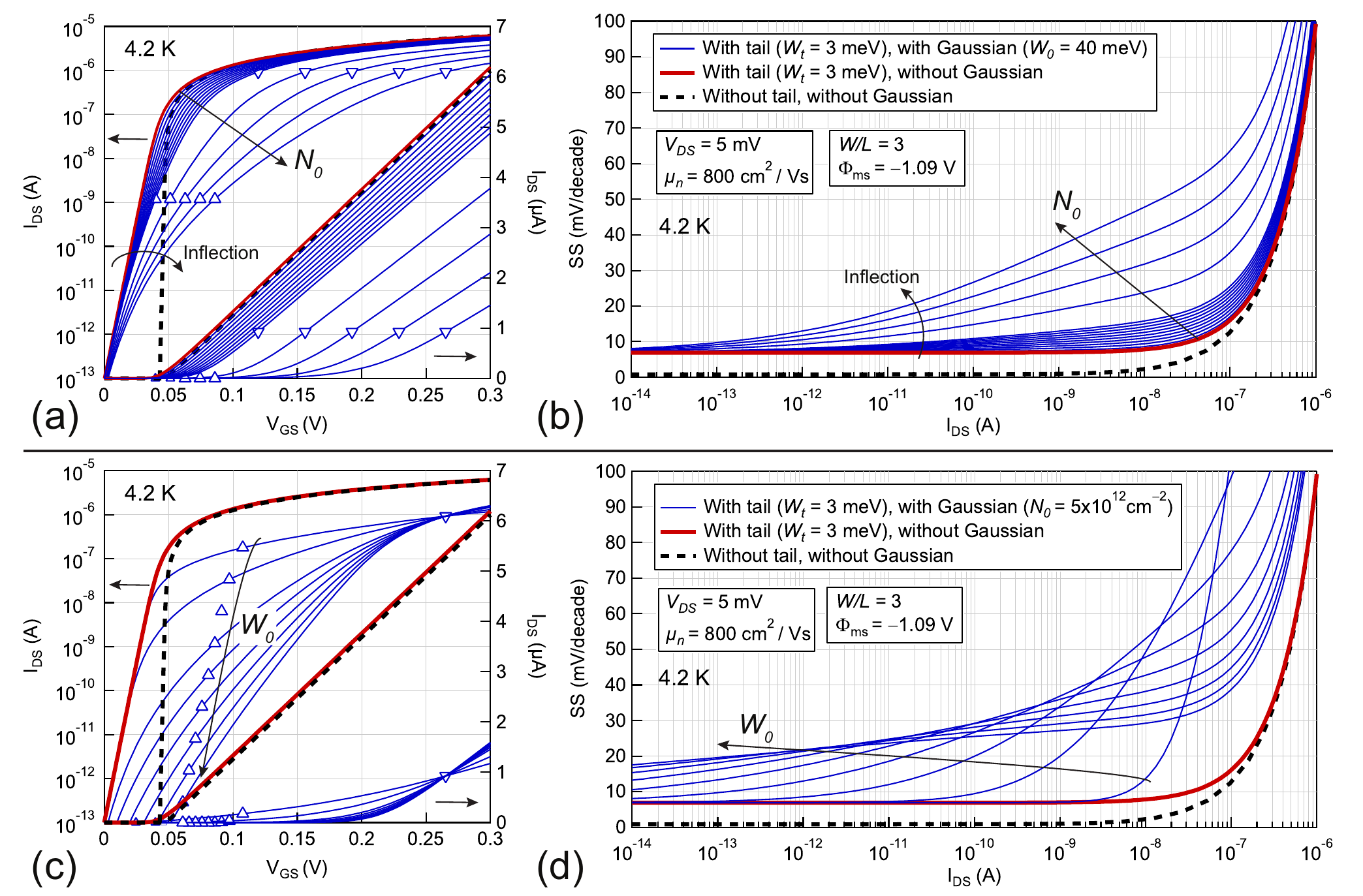}
	\vspace{-0.2cm}
	\caption{Parametric study at 4.2 K (model only) for increasing $N_0$ (top half) and increasing $W_0$ (bottom half). a) The inflection phenomenon (encircled at arbitrary current level) increases when $N_0$ increases from \SI{e11}{\per\centi\meter\squared} to \SI{5e12}{\per\centi\meter\squared} in steps of \SI{1e11}{\per\centi\meter\squared}. This reduces the current and its curvature around moderate inversion. Since the width of the Gaussian is constant, the inflection phenomenon start at the same current level. b) Inflection degrades $SS$. c) and d) Increasing $W_0$ from 10 to \SI{90}{\milli\electronvolt} lowers the current level at which inflection starts.}
	\label{fig:param}
\end{figure*}
\begin{figure}[t]
	\includegraphics[width=0.45\textwidth]{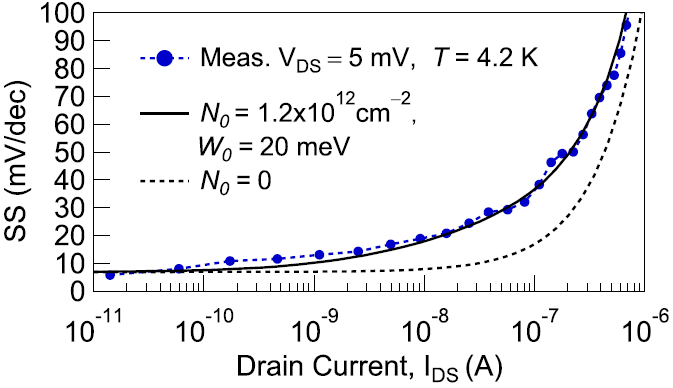}
	\vspace{-0.2cm}
	\caption{Experimental validation. Other physical parameters as in Fig.\,\ref{fig:param}.}
	\label{fig:val}
\end{figure}
\section{\label{sec:model}Results and Discussion}
The model in \cite{bohus} proposes an exponential band broadening to explain $SS$ saturation. This band broadening (or band tail) was attributed to the combined effects of crystalline disorder, strain, residual impurities, surface roughness, etc. To model the inflection within the same framework, it was argued that part of the delocalized states in the exponential band tail is actually localized (end of Section III in \cite{bohus}). Delocalized states are mobile and contribute to the diffusion current. Localized states are trapped and can only modify the gate-voltage dependence and thus the curvature of the diffusion current. Subsequently, a different characteristic decay of the exponential band tail was needed for the localized states to model the change in curvature. Characteristic decays of \SI{3}{\milli\electronvolt} and \SI{10}{\milli\electronvolt} were empirically determined for the delocalized and localized states, respectively [see Fig.\,3(c) in \cite{bohus}]. While appropriate for compact modeling, the latter gives no insight into the relative densities of mobile and trapped states in the band tail. Furthermore, the exponential dependence due to the trapped states was introduced ad hoc, as a multiplicative factor in the expression of the gate-to-source voltage ($V_{GS}$) versus the surface potential ($\psi_s$) : $V_{GS}=m(\psi_s)\psi_s+qn(\psi_s)/C_{ox}$, where $m(\psi_s)$ exponentially depends on $\psi_s$ ($q$ is the electron charge, $n$ is the electron density, and $C_{ox}$ the oxide capacitance). However, a trapped charge density at the silicon/oxide interface would yield an additive term in the continuity of dielectric displacement vectors \cite{sze}. Furthermore, the exponential distribution of localized states proposed in \cite{bohus} diverges for energies above the band edge. In view of the above, we propose to introduce a Gaussian distribution of localized states, as shown in Fig.\,\ref{fig:band}. The Gaussian shape is an improved physical representation of the distribution of localized states, because it does not diverge for energies above the band edge while keeping an exponential tail in the bandgap. All impact of disorder (combined effect of surface roughness, residual impurities, strain, vacancies, etc.) and interface traps at the silicon/oxide interface is now accounted for by this Gaussian. In \cite{bohus}, these disorder effects were assumed to be responsible for the exponential band tail of mobile states (shown by the red line in Fig.\,\ref{fig:band}). That exponential band tail of mobile states is kept for $SS$ saturation, but attributed mainly to intrinsic mechanisms blurring the band edge, e.g., electron-electron interactions, electron-phonon scattering, etc.~\cite{sarang,edl}. The clear distinction between the Gaussian and the exponential band tail allows to estimate the density of localized states at the oxide/silicon interface from the measured curvature, as demonstrated next. 

Without loss of generality, we will assume an $n$-channel, silicon MOSFET with doping concentration $N_A=\SI{e17}{\per\cubic\centi\meter}$, $C_{ox}=\SI{22}{\milli\farad\per\meter\squared}$, and a conduction-band-tail extent of $W_t=\SI{3}{\milli\electronvolt}$ for $SS$ saturation below about \SI{40}{\kelvin} ($W_t$ is typically about 3 to 4 meV\cite{bohus,edl}). Furthermore, we neglect bandgap widening in silicon ($E_g=\SI{1.12}{\electronvolt}$) as well as the impact of substrate freezeout on the Fermi potential [$\Phi_\mathrm{F}=(k_BT/q) \ln(N_A/n_i)$], since these phenomena have little impact on weak and moderate inversion \cite{tedpaper,beckers_jeds}. As shown in Fig.\,\ref{fig:band}, the Gaussian distribution of localized states is centered around the sharp conduction band edge at the silicon/oxide interface ($E_{c,s}$). Note that by inflection point in the Gaussian we mean the strict mathematical meaning of \textquoteleft inflection point\,\textquoteright, marked by an upward triangle, which does not directly correspond to the start of the \textquoteleft inflection phenomenon\textquoteright \, in transfer characteristics or $SS-I_{DS}$ graphs (inflection meaning here \textquoteleft flexing inward\textquoteright). The Gaussian has a width $W_0$ of twice the standard deviation ($\sigma$) and a maximum density of $N_0/(\sigma\sqrt{2\pi})$ at $E_{c,s}$. The localized charge density in the Gaussian can then be expressed as: 
\begin{equation}
Q_0=-q\int^{+\infty}_{-\infty}\frac{N_0}{\sigma\sqrt{2\pi}}\exp\left(\frac{-(E-E_{c,s})^2}{2\sigma^2}\right)f(E)dE.
\label{eq:qgaussianint}
\end{equation}

Equation (\ref{eq:qgaussianint}) is known as the Gauss-Fermi integral, which has no closed-form solution \cite{paasch}. Therefore, we assume that $f(E)$ can be approximated by a Heaviside step function at cryogenic temperatures (only for the $f(E)$ of the localized states). This assumption will limit the validity of the resulting model to the cryogenic regime. We find:
\begin{equation}
Q_0=-q\frac{N_0}{2}\left[\mathrm{erf}\left(\frac{E_{F,n}-E_{c,s}}{(W_0/2)\sqrt{2}}\right)+1\right].
\label{eq:qgaussian}
\end{equation}
From Fig.\,\ref{fig:band}(a), we can find that $E_{F,n}-E_{c,s}=q(\psi_s-\psi_s^*)$, where  $\psi^{*}_s=E_g/(2q)+\Phi_\mathrm{F}+V$ corresponds to a specific amount of band bending for which $E_{c,s}=E_{F,n}$ at the silicon/oxide interface. Here, $V=(E_F-E_{F,n})/q$ is the channel voltage which is not shown in Fig.\,\ref{fig:band}(a). Neglecting the depletion charge density, the continuity of dielectric displacement vectors at the silicon/oxide interface gives $V_{GS}=\psi_s+\Phi_{ms}+qn(\psi_s)/C_{ox}-Q_0/C_{ox}$, where $\Phi_{ms}$ is the metal-semiconductor work function difference. The electron density $n(\psi_s)$ will be considered for the cases with and without the conduction-band tail. Without conduction-band tail and full Fermi-Dirac statistics, $n(\psi_s)$ is given by $n_2(\psi_s)=N_{c}^{\mathrm{2D}}k_BT\ln(1-z^{-1})$, where $z=-\exp\left[-q(\psi_s-\psi_s^*)/(k_BT)\right]$. Here, we assumed for simplicity the constant 2-D conduction-band DOS: $N_c^{2D}=g_vm^*/(\pi\hbar^2)$, where $g_v=2$ is the degeneracy factor, $m^*=0.19\,m_o$ is the effective mass in silicon (assumed temperature independent), and $\hbar$ the reduced Planck's constant. Including the conduction-band tail [red line in Fig.\, \ref{fig:band}(b)] and Fermi-Dirac statistics leads to $n(\psi_s)=n_1+n_2=N_c^{\mathrm{2D}}W_tF_1\left(1,\theta;\theta+1;z\right)+n_2$, where $n_1$ is the electron density in the tail, $F_1$ is the hypergeometric function \cite{edl,abramowitz}, and $\theta\triangleq k_BT/W_t$. Combining the previous, we obtain the following $V_{GS}-\psi_s$ relation: 
\begin{equation}
V_{GS}-\Phi_{ms}=\psi_s+\underbrace{\frac{qn(\psi_s)}{C_{ox}}}_{\mathrm{mobile}}+\underbrace{\frac{qN_0}{2C_{ox}}\left[\mathrm{erf}\left(\frac{q(\psi_s-\psi_s^*)}{(W_0/2)\sqrt{2}}\right)\!+\!1\right]}_{\mathrm{localized}}
\label{eq:vg}
\end{equation}
where the contribution of localized states (disorder, traps) gives an additional RHS term. Since the inflection phenomenon happens at both low and high $V_{DS}$, we assume small $V_{DS}$ here for simplicity, and express the drift-diffusion current as $I_{DS}=(W/L)qn(\psi_s)\mu_n V_{DS}$, where $W/L$ is the aspect ratio of the gate and $\mu_n$ the electron mobility. The trapped charges do not contribute directly to $I_{DS}$, but they can modify the $I_{DS}$ versus $V_{GS}$ characteristic through (\ref{eq:vg}). 

Figure \ref{fig:param} plots $I_{DS}-V_{GS}$ and corresponding $SS-I_{DS}$ graphs at \SI{4.2}{\kelvin} for different cases. First, a standard drift-diffusion current without tail ($n_1=0$) and without Gaussian ($N_0=0$) is plotted with dashed lines in Figs.\ref{fig:param}(a)-(d). Including the tail with $W_t=\SI{3}{\milli\electronvolt}$ degrades $SS$ from $\approx$ 1 mV/dec to $\approx$ 8 mV/dec, as shown by the dark brown lines in Figs.\ref{fig:param}(a)-(d). This saturation of $SS$ is clearly visible in Figs.\,\ref{fig:param}(b) and \ref{fig:param}(d). Then, the localized term in (\ref{eq:vg}) due to the Gaussian is included. At a given $W_0$ [Figs.\,\ref{fig:param}(a) and \ref{fig:param}(b)], the inflection starts at the same current level [around $\SI{e-13}{\ampere}$ and $\SI{e-12}{\ampere}$ in Fig.\,\ref{fig:param}(a) and \ref{fig:param}(b)]. Increasing $N_0$ from \SI{e11}{\per\centi\meter\squared} to \SI{5e12}{\per\centi\meter\squared} (more disorder or traps at the surface), $I_{DS}$ and its slope degrade around moderate inversion (blue lines). The upward triangles correspond to situations when $E_{F,n}$ is at the first inflection point of the Gaussian, and the downward triangles when $E_{F,n}$ is at the maximum of the Gaussian (see Fig.\,\ref{fig:band}). At a given $N_0$ [Figs.\,\ref{fig:param}(c) and \ref{fig:param}(d)], the inflection phenomenon starts at lower current levels when the tail of the Gaussian extends further in the bandgap. In Figs.\ref{fig:param}(c) and \ref{fig:param}(d), $W_0$ increases from 10 to \SI{90}{\milli\electronvolt}. The downward triangle in Fig.\,\ref{fig:param}(c) is now stationary, since $Q_0$ [given by (\ref{eq:qgaussian})] does not depend on $W_0$ at the maximum ($E_{F,n}=E_{c,s}$). Figure \ref{fig:val} shows the experimental validation of the model at \SI{4.2}{\kelvin}. A reasonable density of localized states of about $\SI{e12}{\per\centi\meter\squared}$ and $W_0=\SI{20}{\milli\electronvolt}$ are obtained from fitting the model with the curvature of the measurements. The extracted tail of the Gaussian due to disorder and traps ($\sigma=W_0/2=\SI{10}{\milli\electronvolt}$) is larger than the band tail of mobile states ($W_t=\SI{3}{\milli\electronvolt}$) and agrees well with the electron-spin resonance measurements performed by Jock et al. (see Fig. 3 in \cite{jock}, which shows the decay of the density of trapped electron states below the conduction band edge in a silicon MOSFET).  
\section{Conclusion}
An analytical model is derived for the inflection of cryogenic MOSFET $I_{DS}-V_{GS}$ curves. A Gaussian distribution of localized states is added to the electrostatics. The approximation required to solve the Gauss-Fermi integral limits the scope of the model to cryogenic temperatures. By plotting $SS$ versus $I_{DS}$ from measurements, the width and the maximum of the Gaussian distribution can be determined. These two parameters are important for modeling FETs at cryogenic temperatures in weak and moderate inversion.

\bibliographystyle{IEEEtran}
\bibliography{inflection}
\end{document}